\documentclass[twocolumn]{svjour3}         % twocolumn
\smartqed  % flush right qed marks, e.g. at end of proof
\usepackage{graphicx}
\usepackage{epsfig}
\usepackage{epstopdf}
\begin{document}

\title{Genesis of coexisting itinerant and localized electrons in Iron Pnictides%\thanks{Grants or other notes
}

%\titlerunning{}        % if too long for running head

\author{Luca de' Medici        \and
         Syed R. Hassan \and
       Massimo Capone
 }


\institute{L. de' Medici \at
              Laboratoire de Physique des Solides, Univ. Paris-Sud, CNRS, UMR 8502, 
F-91405 Orsay Cedex, France.  \\
              Tel.: +33-1-6915-4540\\
              Fax: +33-1-6915-6086\\
              \email{demedici@lps.u-psud.fr}           %  \\
              \and
              S.R. Hassan \at
The Institute of Mathematical Sciences
C.I.T. Campus,Taramani, Chennai 600 113, India
           \and
           M. Capone \at
              SMC, CNR-INFM, and Universit\`a di Roma "La Sapienza'',  Piazzale Aldo Moro 2, I-00185 Roma, Italy and  ISC-CNR, Via dei Taurini 19, I-00185 Roma, Italy
}

\date{Received: date / Accepted: date}

\maketitle

\begin{abstract}
We show how the general features of the electronic structure of the Fe-based high-Tc superconductors are a natural setting for a selective localization of the conduction electrons to arise.  Slave-spin and Dynamical mean-field calculations support this picture and allow for a comparison of the magnetic properties with experiments. 
\end{abstract}

\section{Introduction}\label{intro}
Since 1986, each time a new "high-temperature" superconductor has been discovered, it has been a natural temptation to compare the new materials to the cuprates, looking for unifying aspects which could shed light on  the key mechanism for high-temperature superconductivity. This temptation is probably stronger then ever in the case of the iron-based "pnictides", the second family of transition-metal-based superconductors with critical temperature exceeding 50K\cite{discovery}.

The two classes of materials share indeed several features among which we may quote the structure, which is quite similar, with FeAs layers playing a similar role of CuO planes, and the observation that in both cases superconductivity appears doping a magnetically ordered parent compound, and follows a dome behavior as a function of doping. 

An important different lies in the nature of the parent compound.  In the cuprates the half-filled system is an insulator which is usually considered a Mott insulator, testifying the central role of electron correlations (Even if it has been recently proposed that the degree of correlation can be smaller than usually believed\cite{comanac}). The undoped pnictides are instead still metallic, despite the magnetic ordering, suggesting a weaker degree of electronic correlations.
The actual strength of correlation in pnictides is lively debated: while there is no obvious reason to invoke correlation as the key player in the game as it happens in cuprates, it has been proposed that these compounds are actually close to a Mott transition\cite{correlanti}.

Another peculiarity of pnictides  is that, according to density-functional theory calculations, all the five d-bands of iron are relevant  as opposed to the single copper band of the cuprates. Multiple bands crossing the Fermi level are confirmed by angular resolved photoemission. The relevance of multiband effects can solve the discrepancy between different estimates of the role of correlations.
Indeed electron correlation can induce a much richer physics in multiband systems with respect to single band. In particular, different bands can be affected differently by correlations, and eventually one can have an orbital-selective Mott transition (OSMT), in which some bands become insulating and others are metallic.
Signatures of an orbital-selective behavior have been detected by angle-resolved photoemission\cite{arpes}, where all the bands are quite renormalized with respect to density-functional calculations, and the ones lying closest to the Fermi level have a significantly stronger renormalization. An indication of an orbital-selective physics is also provided by the magnetic response, which seems to be constituted by a metallic component (Pauli behavior) and a component associated to local moments (Curie susceptibility), as if two different kind of carriers are present\cite{magnetismo}.

\section{The mechanism: selective localization triggered by band degeneracy lifting}\label{sec:OSMT}

In a recent paper\cite{demedici_3bandOSMT} we have outlined a mechanism for the Orbital-selective Mott transition, which is possibly the most likely to be found in nature, and in particular in the pnictides.

This reposes on the different kinetic energy of  electrons lying in manifolds of bands of unequal degeneracy. In general, even if the bandwidths are equal, the higher the degeneracy, the higher the kinetic energy content. Thus, in case of reasonably decoupled manifolds, intermediate interactions can localize only the less mobile electrons, while the more mobile ones remain itinerant. 

The necessary conditions for this to be realized are thus a suitable splitting of the conduction bands due to the crystal field, and a sizeable Hund's exchange coupling which suppresses the orbital fluctuations and hence decouples the manifolds.

Indeed it is well known that manifolds of degenerate atomic orbitals are often split in energy by the anisotropy of the Crystal field. Distortions of the ionic structure lower the point group symmetry of the crystal and this changes the Madelung energy of the orbitals oriented along different directions. A typical example is the splitting, induced by tetragonal distortions, of the five d-obitals that generate the conduction bands in transition-metal oxydes in two groups, $t_{2g}$ and $e_g$, respectively of three and two bands.

We also know from the study of the multi-orbital Hubbard model that Mott insulating phases can arise at large interaction strength for any integer filling\cite{Lu_gutz_multiorb,rozenberg_multiorb} and it is now clear\cite{gunnarsson_fullerenes,gunnarsson_multiorb,Koch_gap,florens_multiorb} that the critical interaction strength $U_c$ for the Mott transition is in general larger for manifolds of a larger number of bands (e.g. in the SU(N)-orbital Hubbard model it scales with N at large N, while the critical value for the closing of the gap scales with $\sqrt{N}$.) due to the increased kinetic energy of electrons lying in degenerate bands. 
%  
%For a given number of bands the $U_c$  is largest at half-filling and decreases moving away from it if the conduction electrons are split in manifolds of different number of bands this mismatch in $U_c$ can play a role. Indeed if the manifolds were completely decoupled they would undergo a Mott transition at distinct values of U. In \cite{demedici_3bandOSMT} we have shown how this survives the fact that in practice the manifolds are coupled by an interband coulomb repulsion and also a spin-spin exchange interaction (that gives rise to the so-called Hund's rules) and a pair-hopping term, provided the crystal field is large enough to considerably suppress the orbital fluctuations, and the hybridization between them absent or very small.

\begin{figure}[h]
\includegraphics[width=0.5\textwidth]{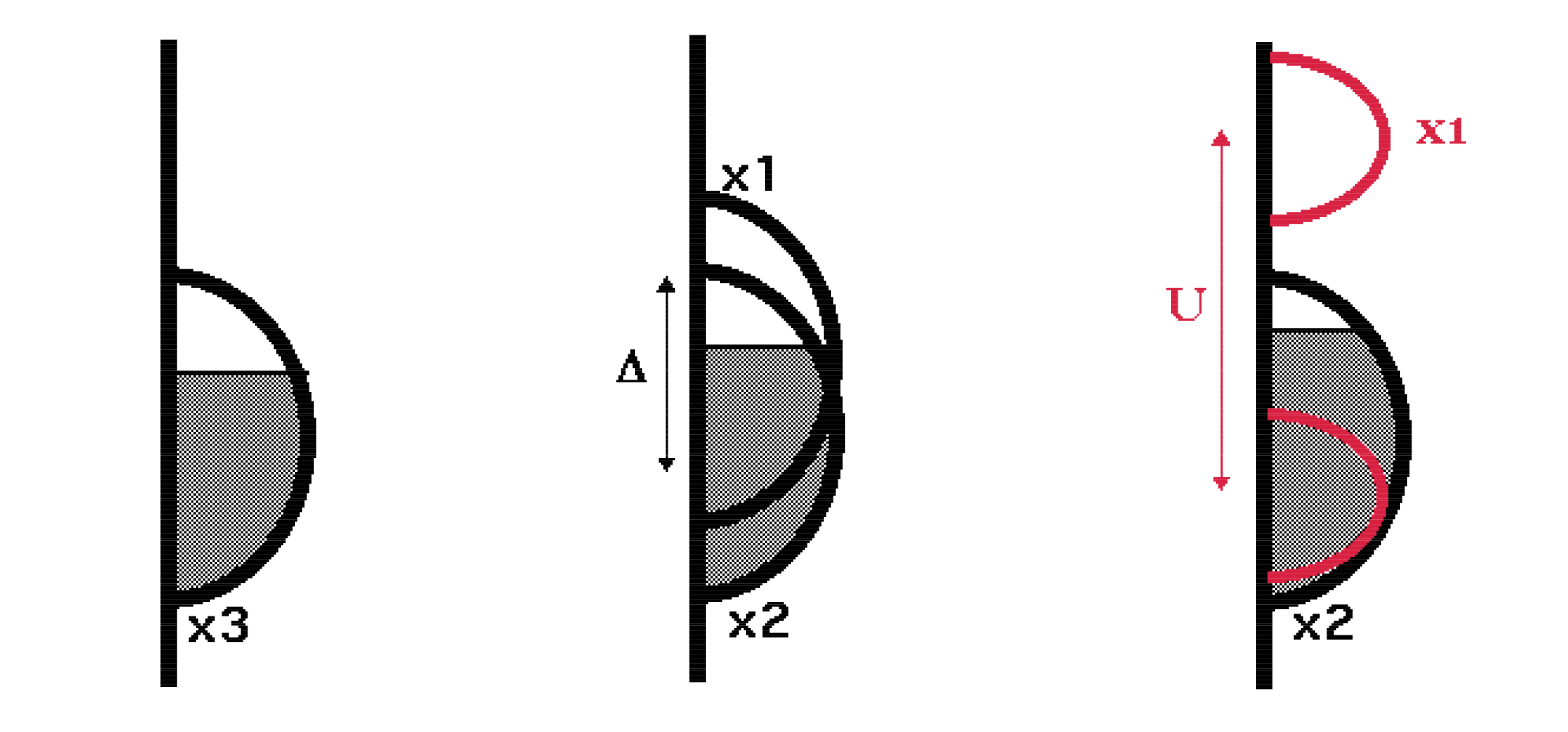}
\caption{Sketch of the band structure and splitting that allows the new OSMT mechanism to apply. The three band-model realization was studied in Ref. \cite{demedici_3bandOSMT}. In this paper we generalize it to the case relevant for the pnictides: 6 electrons in 5 bands. For simplicity we study the case in which one band is lifted and the remaining 4 stay degenerate.}
\label{fig:BandSplit}   
\end{figure}

We analyzed a model in which 4 electrons per site populate three degenerate bands which are split by the crystal field as depicted in Fig. \ref{fig:BandSplit} and found that the OSMT is realized for a large range of the parameter space $U-\Delta$ (interaction strength - crystal field splitting) if the Hund's coupling $J$ is higher than a critical value. 

In practice when $J$ is large enough to suppress the orbital fluctuations, a gap due to $U$ can open in the lifted band, such that the chemical potential falls within it and this band is half-filled. This, and the related localization can happen for the whole range of the crystal field that shifts the gapped band until the chemical potential hits one of the borders of the gap.

\section{Towards a realistic model for iron pnictides: 6 electrons in 5 bands}

Here we generalize the study performed in Ref. \cite{demedici_3bandOSMT} by investigating the realization of the OSMT by the same mechanism, within a more appropriate (albeit still rather caricatured) model for Iron Pnictides.

Indeed the case relevant for Pnictides\cite{HauleShim_FeAs,Haule_FeAs}, 6 electrons in 5 bands, is even more favorable than the previous one of 4 electrons in 3 bands, since it is closer to global half-filling. Thus an even smaller crystal field splitting can drive the top band to half-filling and to localization. 

The crystal field in these compounds is estimated to split the orbitals by putting the $xy$ 130 meV above the degenerate $xy$ and $yz$, that are in turn 160meV above the $3z^2-r^2$ and 220meV above the $x^2-y^2$. The Hund's coupling is estimated of the order of $0.35\sim0.7$ eV while the bandwidth is $W\sim 3eV$ and $U\sim4eV$. 

This situation is ideal for our mechanism to apply, since the upper band is likely to be driven towards half-filling by the crystal field and the $U/W$ value is just about what is needed for a single band to be localized by correlations. The top band is relatively close in energy to the doublet but this, even if its population reduced from $6/5$ by the crystal field, has larger critical U because of its degeneracy, certainly above the realistic value of the Coulomb repulsion in these materials. The other bands are pushed at still lower energies and will probably be populated by more than $6/5$ particles.

\begin{figure}[h]
\includegraphics[width=0.5\textwidth]{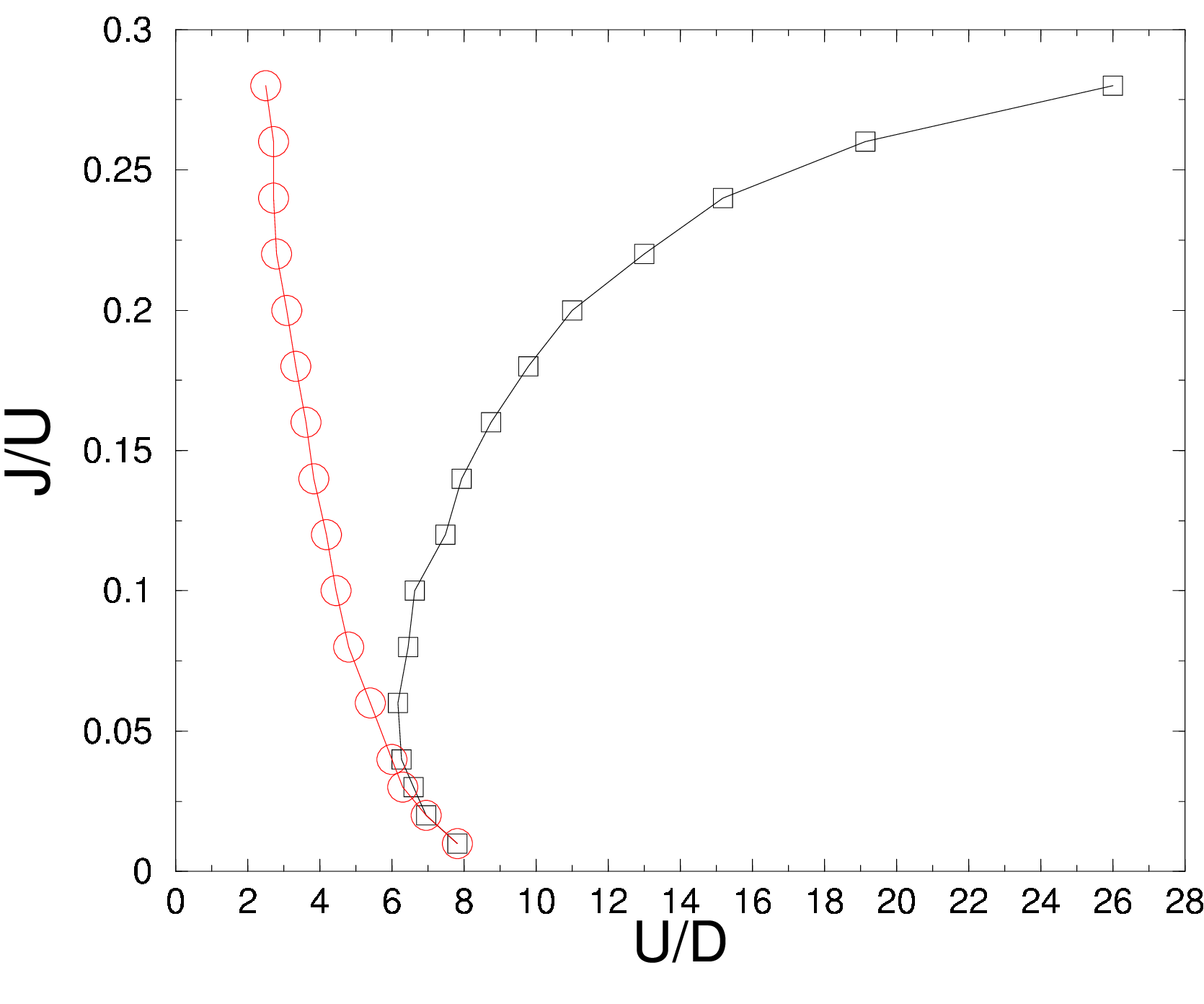}
\caption{Phase diagram (Slave-spin mean-field), as a function of U and J, of a 5-band model with four bands degenerate and one lifted by the crystal field so to be always half-filled. The other bands host the remaining 5 electrons. $D$ is the half-bandwidth.}
\label{fig:PhD5bands}   
\end{figure}

We did not enter in the realistic calculation with all the details of the crystal field splitting, this is left for further work. We only show here (figure \ref{fig:PhD5bands}) the phase diagram of the system with four degenerate bands and one lifted by the crystal field in the (1,5) populated phase. It is easy to see that as predicted the OSMP is more easy to trigger than in the previous case (Fig. 1 of Ref. \cite{demedici_3bandOSMT}) due to the increased proximity to half-filling. Indeed a smaller critical J is found and also a smaller crystal field is enough to enter the OSMP.

We also notice that within our naive model and crude approximation the values estimated for the local interaction in the pnictides ($U\sim 2.7D,  J/U\sim 0.1\div 0.2$) fall very close to the border of the OSMT. In this regime, even if the zero-temperature transition has not happened yet, the renormalization of the electronic properties differs very strongly between the bands, the almost localized one having a very low coherence temperature compared to the others. Thus even if pnictides are not strictly in the zone of the phase diagram showing a selectively localized ground state (and even this has to be confirmed by more accurate methods in more realistic models), most probably they lie in a regime of selective localization at finite temperature.

\section{Magnetic properties and comparison with experiments}

A key quantity for the individuation of the OSMT in materials is the local spin susceptibility $\chi_{loc}$. Indeed while the itinerant electrons will dominate the transport properties, the presence of a localized component will show up, thanks to the magnetic moments that form.
 
\begin{figure}[h]
 \includegraphics[width=0.5\textwidth]{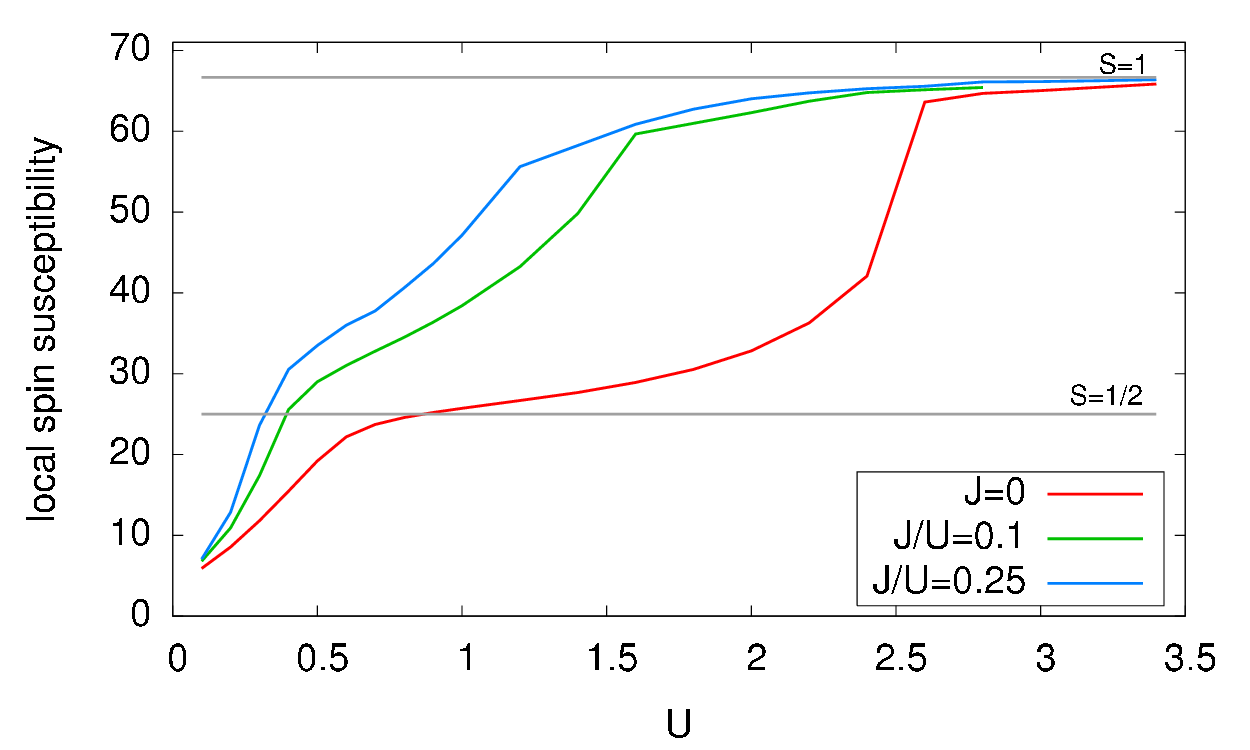}
\caption{Local magnetic susceptibility in the "classic" model for OSMT (2 bands of unequal bandwidth), as a function of U/D, for different strength of the Hund's coupling. It is easily seen that when $J>0$ the local magnetic moment that develops in the selectively localized phase is bigger than the expected $S=1/2$, and grows with J.}
\label{fig:susc} 
\end{figure}

Dynamical mean-field theory allows a reliable calculation of this quantity. However the 5-band model under examination here requires a tremendous numerical effort to be tackled reliably by this technique. Here we just want to show some general features of the local magnetic susceptibility of a selectively localized phase, namely the dependence of the size of the local moment on the Hund's coupling J. In Fig.\ref{fig:susc} we show this dependence in a much simpler model, paradigmatic for the OSMT, of two bands with different bandwidth ($W_2=0.15W_1$), in which the narrower one gets localized by correlations while the wider one remains itinerant.

When a band gets localized the susceptibility goes from a Pauli-like to a Curie-like behaviour indicating the formation of free moments. 
As expected for an OSMT in Fig. \ref{fig:susc} one sees a two stage saturation of the $\chi_{loc}$ signaling the partial localization followed by the complete Mott insulating state with magnetic moments corresponding to the high-spin atomic state. 

One could naively expects that the value at which the $\chi_{loc}$ saturates upon entering the selective phase be the one for a local $S=1/2$ moment.
However the free moments in the OSMP are coupled to the itinerant electrons by the exchange interaction. As a result we see that in this phase the magnetic susceptibility shows a Curie component but the value of the magnetic moment of the localized electrons is enhanced by the presence of the itinerant electrons. Indeed the Hund's coupling ferromagnetic correlation between the on-site spin in the different orbitals induces a local moment in the itinerant bands. The value of the total effective moment is thus larger than the simple one of the localized component and grows with $J$.

Indeed this can be shown by very simple arguments using the effective model for the OSMP introduced in \cite{Biermann_nfl}. In this simplified picture of the OSMP one can show that the local magnetic susceptibility has the form

\begin{equation}
\chi_{loc}=\chi_{it}+\frac{(M_{it}[J/2] + \frac{1}{2})^2}{T},
\end{equation}
where $\chi_{it}$ is the local susceptibility of the itinerant part (still Pauli-like) and $M_{it}(H)$ is the magnetic moment induced by a local field H in the itinerant component. 

A detailed comparison between experimental results for the magnetic response and a more realistic calculation based on the combination of many-body approaches and realistic bandstructure is the natural extension of the present analysis. It is worth noting that through a similar comparison one could obtain an accurate estimate for an otherwise elusive quantity, the Hund's interaction $J$.

In this paper we have discussed the possibility of an orbital-selective Mott transition in a five-band model which is a backbone description of the electronic structure of the pnictides. We found a realization of the transition proposed in Ref. \cite{demedici_3bandOSMT}, where the OSMT is determined by the lifting of the orbital degeneracy due to a (reasonably small) crystal field splitting. 
The relevance of band-selective properties in multiband superconductivity has been indeed proposed in the context of the cuprates\cite{multicuprates} and pnictides\cite{caivano}

\begin{acknowledgements}
We acknowledge financial support of MIUR PRIN 2007 Prot. 2007FW3MJX003
\end{acknowledgements}

\bibliographystyle{spphys}       % APS-like style for physics
\bibliography{Bib/publdm,Bib/bibldm}   % name your BibTeX data base

\begin{thebibliography}{10}
\providecommand{\url}[1]{{#1}}
\providecommand{\urlprefix}{URL }
\expandafter\ifx\csname urlstyle\endcsname\relax
  \providecommand{\doi}[1]{DOI \discretionary{}{}{}#1}\else
  \providecommand{\doi}{DOI \discretionary{}{}{}\begingroup
  \urlstyle{rm}\Url}\fi

\bibitem{discovery}Y.~Kamihara {\sl et al.}, J. Am Chem. Soc. {\bf 128}, 10012 ( 2006); {\it ibid.} {\bf 130}, 3296 (2008)

\bibitem{comanac} A,~Comanac, L.~de' Medici, M.~Capone, and A.~J.~Millis, Nat. Phys. {\bf 4}, 287 (2008)

\bibitem{correlanti} Q.~Si, E.~Abrahams, J.~Dai, J.~X.~Zhu, arXiv:0901.4112v1

\bibitem{arpes} Y. Sekiba, T. Sato, K. Nakayama, K. Terashima, P. Richard, J. H. Bowen, H. Ding, Y.-M. Xu, L. J. Li, G. H. Cao, Z.-A. Xu, T. Takahashi, arXiv:0812.4111; H. Ding, K. Nakayama, P. Richard, S. Souma, T. Sato, T. Takahashi, M. Neupane, Y.-M. Xu, Z.-H. Pan, A.V. Federov, Z. Wang, X. Dai, Z. Fang, G.F. Chen, J.L. Luo, N.L. Wang, arXiv:0812.0534

\bibitem{magnetismo} T.~Nomura, S.~W.~Kim, Y.~Kamihara, M.~Hirano, P.~V.~Sushko, K.~Kato, M.~Takata, A.~L.~Shluger, H.~Hosono, Supercond. Sci. Technol. {\bf 21}, 125028  (2008)

\bibitem{demedici_3bandOSMT}
L.~de' Medici, S.R.Hassan, M.~Capone, X.~Dai,   (2008).
\newblock Phys. Rev. Lett. \textbf{102}, 126401 (2009)

\bibitem{Lu_gutz_multiorb}
P.~Lu, Phys. Rev. B \textbf{49}, 5687 (1994)

\bibitem{rozenberg_multiorb}
M.J. Rozenberg, Phys. Rev. B \textbf{55}, R4855 (1997)

\bibitem{gunnarsson_fullerenes}
O.~Gunnarsson, E.~Koch, R.~Martin, Phys. Rev. B \textbf{54}, R11026 (1996)

\bibitem{gunnarsson_multiorb}
O.~Gunnarsson, E.~Koch, R.~Martin, Phys. Rev. B \textbf{56}, 1146 (1997)

\bibitem{Koch_gap}
E.~Koch, O.~Gunnarsson, R.M. Martin, Phys. Rev. B \textbf{60}, 15714 (1999)

\bibitem{florens_multiorb}
S.~Florens, A.~Georges, G.~Kotliar, O.~Parcollet, Phys. Rev. B \textbf{66},
  205102 (2002)

\bibitem{HauleShim_FeAs}
K.~Haule, J.~Shim, G.~Kotliar, Phys. Rev. Lett. \textbf{100}, 226402 (2008)

\bibitem{Haule_FeAs}
K.~Haule, G.~Kotliar,  arXiv:0805.0722

\bibitem{georges_RMP_dmft}
A.~{Georges}, G.~{Kotliar}, W.~{Krauth}, M.J. {Rozenberg}, Rev. Mod.  Phys. \textbf{68}, 13 (1996)

\bibitem{pruschke_Hund}
T.Pruschke, R.~Bulla, Eur. Phys. J. B \textbf{44}, 217 (2005)

\bibitem{Biermann_nfl}
S.~Biermann, L.~de' Medici, A.~Georges, Phys. Rev. Lett. \textbf{95}, 206401  (2005)

\bibitem{multicuprates} N.~Kristoffel,  P.~Rubin and T.~\"Ord,  J. Phys.: Conf. Ser. \textbf{108} 012034; F.~V.~Kusmartsev and M.~Saarela, Supercond. Sci. Technol. \textbf{22}, 014008 (2009)

\bibitem{caivano} R. Caivano {\sl et. al.}, Supercond. Sci. Technol. \textbf{22}, 014004  (2009)

\end{thebibliography}

\end{document}